\documentclass[article]{aa}  

\usepackage{graphicx}
\usepackage{txfonts}
\usepackage{lscape}
\usepackage{graphicx}
\usepackage{subcaption}
\usepackage{hyperref}

\begin{document} 

\title{Polarized, variable radio emission from the scallop-shell binary system DG CVn}
\authorrunning{S. Kaur et al.}
\titlerunning{DG CVn polarized radio emission}
\author{Simranpreet Kaur \inst{1,2}\thanks{E-mail: kaur@ice.csic.es}, Daniele Vigan\`o\inst{1,2,3}, Jackie Villadsen\inst{4}, Josep Miquel Girart\inst{1,2},  V\'ictor J.S. B\'ejar\inst{5,6}, Yutong Shan\inst{7,8}, Luke Bouma\inst{9}, Ekaterina Ilin\inst{10}, \`Oscar Morata\inst{1,2},  Miguel P\'erez-Torres\inst{11,12,13}, Etienne Bonnassieux \inst{11}, Jorge R. Gherson \inst{4} 
}

\institute{Institut de Ci\`encies de I'Espai (ICE-CSIC), Campus UAB, Carrer de Can Magrans s/n, 08193 Cerdanyola del Vallès, Catalonia, Spain
\and
Institut d’Estudis Espacials de Catalunya (IEEC), 08860 Castelldefels, Barcelona, Catalonia, Spain
\and
Institute of Applied Computing \& Community Code (IAC3), University of the Balearic Islands, Palma, 07122, Spain
\and
Department of Physics \& Astronomy, Bucknell University,
Lewisburg, PA, USA.
\and
Instituto de Astrofísica de Canarias (IAC), 38205 La Laguna, Tenerife, Spain
\and
Departamento de Astrofísica, Universidad de La Laguna (ULL), 38206 La Laguna, Tenerife, Spain
\and
Centre for Planetary Habitability, Department of Geosciences, University of Oslo, Sem Saelands vei 2b, 0315 Oslo, Norway 
\and
Institut f\"ur Astrophysik, Georg-August-Universit\"at, Friedrich-Hund-Platz 1, 37077 G\"ottingen, Germany
%\and
%Department of Astronomy, California Institute of Technology, Pasadena, CA 91125, USA
\and
Observatories of the Carnegie Institution for Science, Pasadena, CA 91101, USA
\and
ASTRON, Netherlands Institute for Radio Astronomy, Oude Hoogeveensedijk 4, Dwingeloo, 7991 PD, The Netherlands
\and
Instituto de Astrofísica de Andalucía (IAA-CSIC), Glorieta de la Astronomía s/n, E-18008, Granada, Spain
\and
Center for Astroparticles and High Energy Physics (CAPA), Universidad de Zaragoza, E-50009 Zaragoza, Spain
\and
School of Sciences, European University Cyprus, Diogenes street, Engomi, 1516 Nicosia, Cyprus
}
\date{Received xx xx 2025 / Accepted xx xx 2025 }

\abstract
{DG CVn is an eruptive variable star and represents the closest member of the known sample of complex periodic variables, or scallop-shell stars. Over the years, this M dwarf binary system has shown significant flaring activity at a wide range of frequencies. Here, we present a detailed analysis of $\sim 14$ hours of radio observations of this stellar system, taken with the Karl G.~Jansky Very Large Array at band L, centered at 1.5 GHz. In both $7$-hour long observations we have found a quiescent, weakly polarized component, that could be ascribable to the incoherent, gyro-synchrotron emission coming from the magnetosphere surrounding one or both stars, along with multiple $\sim90\%$ right-circularly polarized bursts, some of which last for a few minutes, while others being longer, $\gtrsim$ 30 minutes. Some of these bursts show a drift in frequency and time, possibly caused due to beaming effects or the motion of the plasma responsible for the emission. We assess the possible modulation of burst frequency with the primary and secondary periods, and discuss the properties of these bursts, favoring electron cyclotron maser over plasma emission as the likely underlying mechanism. We compare DG CVn's dynamic spectrum to other young M dwarfs and find many similarities. A dedicated proper radio/optical simultaneous follow-up is needed to monitor the long-term variability, increase the statistics of bursts, in order to test whether the co-rotating absorbers detected in optical can drive the observed radio emission, and whether the occurrence of radio bursts correlates with the rotational phase of either stars.}


\maketitle
%
%-------------------------------------------------------------------

\section{Introduction}\label{sec:intro}

M dwarfs are the most abundant stars in the Galaxy \citep{Chabrier_Abundance, Henry_06} and are frequently characterized by their strong magnetic fields, typically hundreds to kilo-gauss \citep{reiners22}, along with recurrent flaring activity observed across the electromagnetic spectrum from radio waves to X-rays \citep{villadsen19,Joseph_TESS_EROSITA}. Several surveys indicate that mid-to-late M dwarfs flare with remarkable frequency, with over $40\%$ of stars of spectral type M4 or later exhibiting detectable flares, which is over an order of magnitude higher than what is observed in Sun-like stars \citep{Gunther_TESS1_2, yiu24}.

The recent advancements in radio astronomy and several wide radio surveys have unveiled a growing, yet still very small, fraction of main sequence stars that are radio loud \citep{yiu24,driessen24}.
Most of them are M dwarfs, mostly chromospherically active, and/or interacting binary systems. In M dwarfs, radio emission can generally be categorized into two categories. The first one is seen as an apparently steady (quiescent) component, which is relatively weak, often non- or weakly polarized, broad-band, similar to solar microwave flares \citep{dulk85, Alinssandrakis_1986, yiu24}; it can typically be ascribed to incoherent gyro-synchrotron radiation.  
The second class are transient events, which are the intense flare-associated bursts characterized by narrow bandwidths, high brightness temperatures (often $T_B \gg 10^{12}$ K) and strong circular polarization \citep{dulk85, Melrose_dulk_1982, Melrose_2017}. These are thought to be produced by two possible coherent emission mechanisms: plasma emission and electron cyclotron maser (ECM). The latter may be driven by aurora-like current systems in stellar magnetospheres similar to what is seen in ultra-cool dwarfs \citep{mclean11, Kaur_2024b, bloot24}, and is responsible for Jupiter’s intense decametric radio emissions \citep{zarka99, zarka04}. 

Radio bursts are of specific interest because they offer a unique window into the underlying magnetic fields and particle acceleration mechanisms, providing deeper insights into the energetic processes that drive their dynamic stellar environments \citep{mclean11, villadsen19, bloot24, Kaur_2024b,Kavanagh24}. Notably, some M dwarfs are known to show periodic, pulse-like radio flares which are related to the star’s rotation period. Two striking examples of this behavior are UV Cet \citep{Zic_19_UV_Ceti} and AU Mic \citep{bloot24}, both with very high circular polarization. AD Leo is another nearby flare star, which emits with variability over different timescales, as short as milliseconds \citep{zhang23}.

In this paper, we study DG CVn, also cataloged as TIC 368129164 in TESS, and GJ 3789. Its SIMBAD coordinates are $RA_{\rm J2000}=13^{\rm h}31^{\rm m}46.6^{s}$ and $\delta_{\rm J2000}=+29^{\circ}16'36.7''$, and it exhibits a high proper motion of $\mu_{RA}=-242.8\pm0.5$ mas yr$^{-1}$ and $\mu_\delta=-149.8\pm0.3$ mas yr$^{-1}$ \citep{gaia21}. Further, DG CVn forms a visual binary of two M4 V components with a projected separation of $\sim$0.2$''$ (i.e., a few AU). It is a well-known eruptive variable \citep{Beuzit_2004, cortes17, helfand99}, and the Swift satellite detected a gamma-ray super flare in 2014 \citep{osten16}, followed by two bright ($\sim$100 mJy) radio flares at 15 GHz, among the most luminous incoherent radio flares ever observed from a red dwarf star, before returning to a quiescent level of 2–3 mJy after $\sim$ 4 days \citep{fender15}. The target has also been detected at lower frequencies in the LOFAR, VLASS and ASKAP surveys \citep{yiu24,driessen24} (see Sect. \ref{sec:long term radio light curve}), but those snapshot observations did not reveal whether the emission is burst-like or quiescent.

At a reported distance of $d=18$ pc, DG CVn is also the closest confirmed scallop-shell star (SSS), also known as complex periodic variables. These are a newly-identified class of objects characterized by multiple oddly-shaped dips in their optical light curve. This suggests the presence of opaque co-rotating material that could either be gas from the star trapped in huge prominences, or dust, coming from the disk debris or an out-gassing rocky planet \citep{bouma24, stauffer17, zhan19, koen21, waugh22}. Like all SSSs, DG CVn is a young system, with several age diagnostics pointing at an age between 40–150 Myr, depending on its possible association with the Carina–Columba complex \citep{Riedel} or the AB Doradus moving group \citep{Bell_2015}. Its fast rotation indicates the system has an age probably younger than 150 Myr, too \citep{Engle_guinan_2023}. In the TESS-based catalog of SSSs by \citet{bouma24}, DG CVn shows photometric periods of a few hours: $P_{1} = 6.44$ h and $P_{2}=2.60$ h, possibly interpreted as the rotational period of the two stars \citep{bouma24}. Such periodicity is consistently seen in different TESS sectors, meaning that it is stable over several years at least, although the morphology of the dips substantially change (see App.~\ref{app:TESS} for the TESS light curves of DG CVn and a short discussion).

Radio observations can cast light into the enigma by distinguishing if the radio emission is a counterpart of the stochastic chromospheric activity commonly seen in young, fast-rotating M dwarfs, or auroral emission either associated to fast rotation, or caused by magnetic interaction with the occulting material. For instance, centrifugal breakout of co-rotating clumps has been considered as possible mechanisms even for non-SSS stars such as UV Cet \citep{bastian22}. A cross-matching of the most complete SSS sample \citep{bouma24} with the recent radio-star catalogs based on VLASS, LoTSS and ASKAP-RACS surveys \citep{yiu24, driessen24} suggests that there are currently few more known radio-loud SSSs. The (biased) observed radio occurrence of SSS is in line with the sample of young M dwarfs, and is likely a combination of sporadicity of transient radio bursts, limited observational coverage, and low quiescent radio luminosity. In this context, the only two SSSs with dedicated, long radio observations, as far as we know, are: DG CVn, presented in this study, and 2MASS J05082729$-$2101444 (J0508-21) with the upgraded Giant Metrewave Radio Telescope at 575-720 MHz \citep{Kaur_2024b}, who proposed evidence of coherent radio auroral emission and discussed whether the radio and optical behavior could be ultimately powered by the same orbiting material. In this context, exploring the radio behavior of SSSs like J0508-21 and DG CVn, along with well-coordinated photometric campaigns, could help shed light on the physics responsible for this poorly-understood type of young low-mass stellar systems.

In this article, we present a detailed analysis of 14 hours of archival (2018) observations of DG CVn with Karl G.~Jansky Very Large Array (VLA) at band L (1-2 GHz). In Sect.~\ref{sec:observations}, we describe the VLA observations and the data analysis. In Sect.~\ref{sec:results}, we present the results, discussing the dynamic spectrum morphology and possible periodic variability, and compare the results to other M dwarfs in Sect.~\ref{sec:discussion}.

\begin{figure}
\centering
\includegraphics[width=0.48\textwidth]{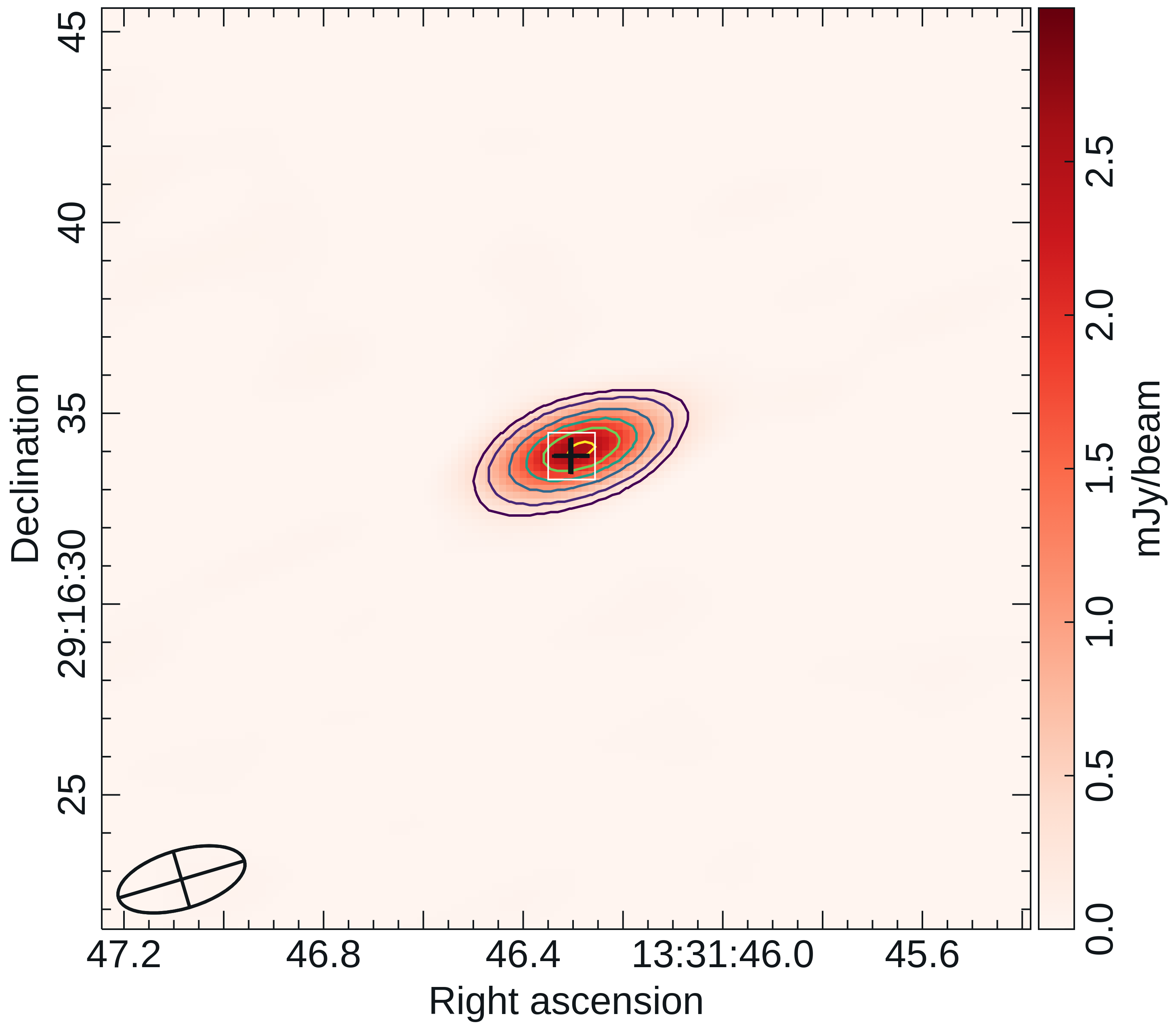}
\caption{Combined map for the January and February observations, centered at the emission peak, which shows an offset from the SIMBAD position which is compatible with the reported proper motion of DG CVn. The synthesized beam is shown in the lower-left corner and the cross (+) marks the proper motion corrected position. The color scale represents the Stokes I flux density, while the contours indicate the Stokes V signal, corresponding to levels of 5, 10, 20, 30, 40 and 50 $\sigma^V_{rms}$ (where $\sigma^V_{rms}$ is $\sim$ 11.2~$\mu$Jy/beam).}
\label{fig:carta_map}
\end{figure}

\begin{table*}[ht]
\caption{VLA band L observations. The fluxes were calculated using the task {\tt imfit} and performing a Gaussian fitting. The errors in the peak and integrated flux represent the statistical standard deviation from the Gaussian fitting.}
\label{tab:flux_statistics}
\centering          
\begin{tabular}{c c c c c c c c} 
\hline\hline
&   & \multicolumn{3}{c}{Stokes I} & \multicolumn{3}{c}{Stokes V } \\    % table heading 
\cline{3-8}
Date & Time on-source & Peak & Integrated   & $\sigma_{I}$   & Peak         & Integrated & $\sigma_{V}$ \\
(dd/mm/yy) & (hours) & (mJy/beam) & (mJy) & (mJy/beam) & (mJy/beam) & (mJy) & (mJy/beam)\\
\hline                    
21/01/18   & 6.1 &$3.77\pm0.06$ &$3.58\pm0.10$& 0.05&$1.49\pm0.05$&$1.43\pm0.08$&0.04\\ 
20/02/18   & 5.7 &$2.79\pm0.02$ &$2.86\pm0.04$& 0.02 &$0.54\pm0.01$& $0.53\pm0.02$&0.01\\ 
Combined    & 11.8 & $3.25\pm0.04$ & $3.38\pm0.07$ &0.03 & $1.02 \pm 0.03$ & $1.18 \pm 0.05$ & 0.03\\ 
\hline                  
\end{tabular}
{\tablefoot{The columns give the date, peak, and integrated fluxes of the source and the rms of the calibrated clean images, for Stokes I and V, for each observation and their combined image.}
}
\end{table*}

\section{VLA Observations and data reduction}\label{sec:observations}

The target DG CVn was observed at band L (1-2 GHz), as part of the VLA proposal 17B-370 (PI: J. Villadsen), on 2018, January 21$^{\rm st}$ (configuration B) and February 20$^{\rm th}$ (BnA), covering $\sim 6$ hours on the target in both the epochs (the second observation was split by a gap of $\sim 1$ hour). For both observations, 3C286 was observed as the standard flux calibrator, while J1330+2509 served as the gain calibrator. The observations were performed in continuum standard mode with 16 spectral windows, each having a bandwidth of 64 MHz, and 64 (1-MHz wide) channels. 

The entire dataset underwent processing via Common Astronomy Software Applications \citep[CASA;][]{CASA}. The calibration of the data adheres to the standard CASA VLA pipeline. For VLA band L, both Stokes V and I are formed from the parallel-hand correlation products, RR and LL, with Stokes I = (LL+RR)/2 and Stokes V = (RR-LL)/2. The observations did not include polarization calibrators. Therefore, we did not perform linear polarization measurements from the crosshand products, RL and LR. After an additional round of manual flagging to correct for the remaining bad data, the maps were constructed with the CASA task  {\tt tclean} by using the {\tt standard} gridding algorithm and employing a {\tt natural}
weighting scheme across all image planes. Further, we used the Cube Analysis and Rendering Tool for Astronomy  \citep[CARTA; ][]{carta_1,carta_2} for image visualization.

We used CASA task {\tt imfit} to extract the flux statistics from the image plane for the LL and RR correlations, and the Stokes I and V parameters. We summarize these flux statistics in Table \ref{tab:flux_statistics}. We applied the same approach to examine the time variability in flux from the image plane by making clean maps for every 5-minutes scan on the target.

To build the dynamic spectrum, we shifted the phase center to the position of our source, averaged the data over all baselines, and made use of CASA task {\tt visstat} to compute the statistics for the real part of the amplitude from the visibility plane. It computes the statistics for the RR and LL correlators separately. From these statistics, we evaluated Stokes V component as (RR-LL)/2, following the standard IAU convention. In a dataset where there are no bright background sources, the real part of the amplitude gives a fair estimate of the flux at the phase center, while the imaginary component of the amplitude reflects the noise level (see e.g., \citealt{villadsen19} for more discussion). The analysis of dynamic spectrum allowed us to infer the flux variability for even shorter time bins and for every frequency channel, allowing us to  uncover the presence of relatively bright bursts. As we are particularly interested in coherent, highly circularly polarized bursts, we did not perform additional subtraction of the background sources, as there were no nearby sources brighter than the target in Stokes V in the field of view which could have possibly contaminated the emission at the phase center.

\section{Results}\label{sec:results}

\begin{figure}[ht!]
\centering
\includegraphics[width=0.47\textwidth]{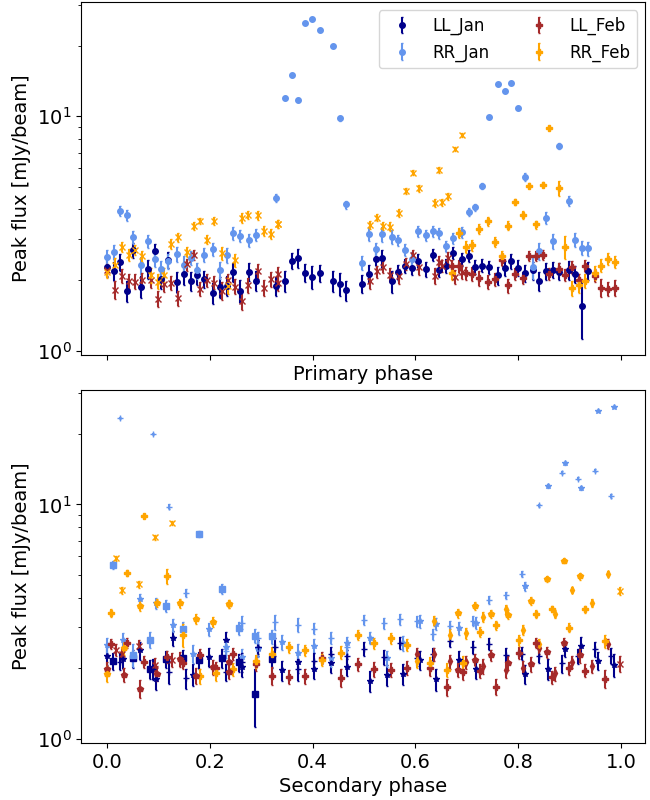}
\caption{Phase-aligned light curves for LL and RR correlations using the primary period $P_1 = 6.44$ h (top) and the secondary period $P_2 = 2.60$ h (bottom), as reported in \cite{bouma24}. The flux represents the peak flux values in logarithmic scale extracted from the image planes of each 5-minutes scan , using the CASA task {\tt imfit}. Here, phase $\phi_{1,2}^{0}$ (ie. starting phase for both periods) corresponds to the beginning of the first target scan of January observation (07:17:00 UT). Different symbols mark the different cycles covered during each observation. Note that the accuracy on the period is enough to guarantee phase linkage after $\sim$1 month,  with an accuracy of $\lesssim$ few $\%$, estimated by eq.~(9) of \cite{bouma24}.}
\label{fig:time_series}
\end{figure}

\begin{figure*}[h!]
  \centering
  \begin{subfigure}[t]{0.95\linewidth}
\centering
\includegraphics[width=\linewidth]{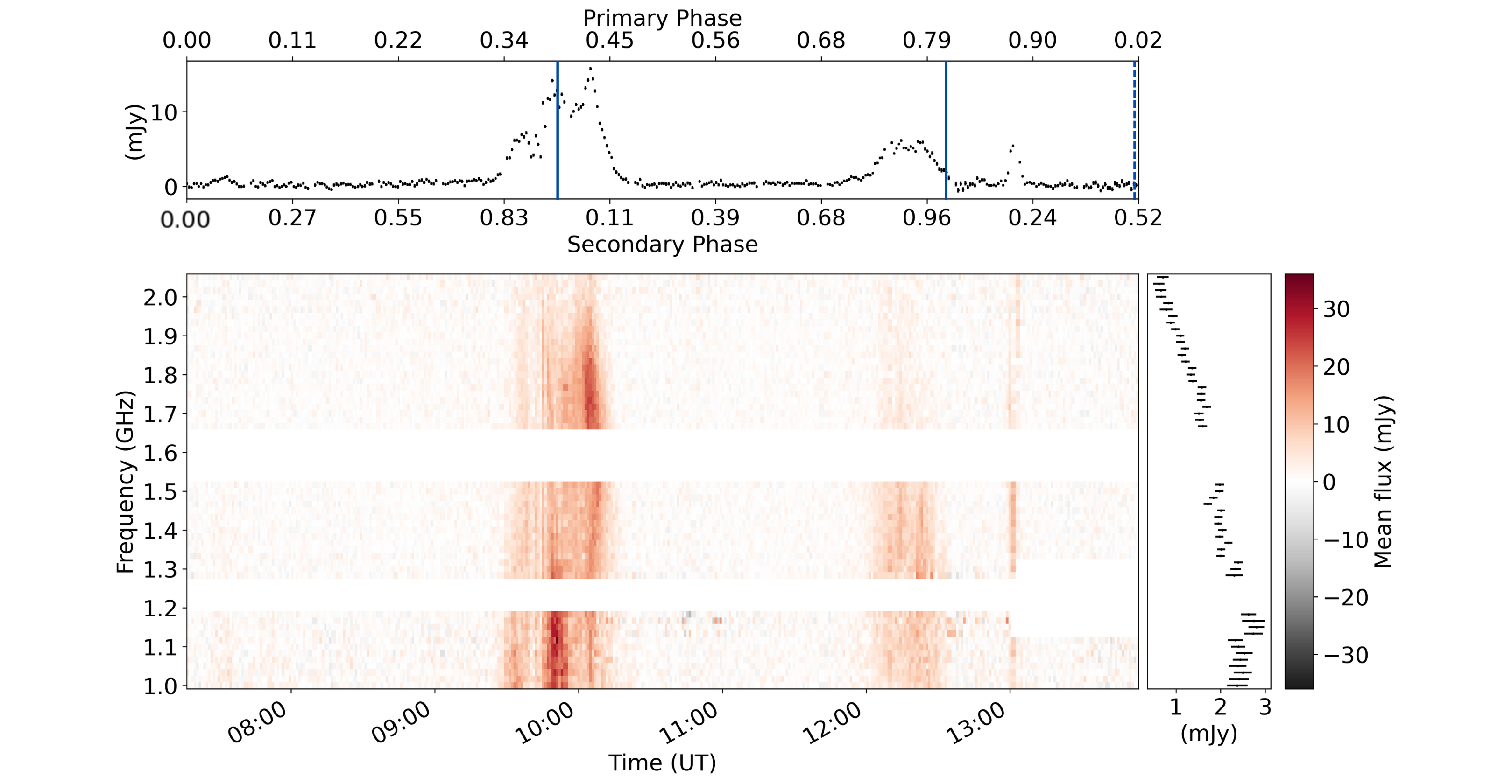}
  \end{subfigure}
   \\[0.5cm]
  \begin{subfigure}[t]{0.95\linewidth}
\centering
\includegraphics[width=\linewidth]{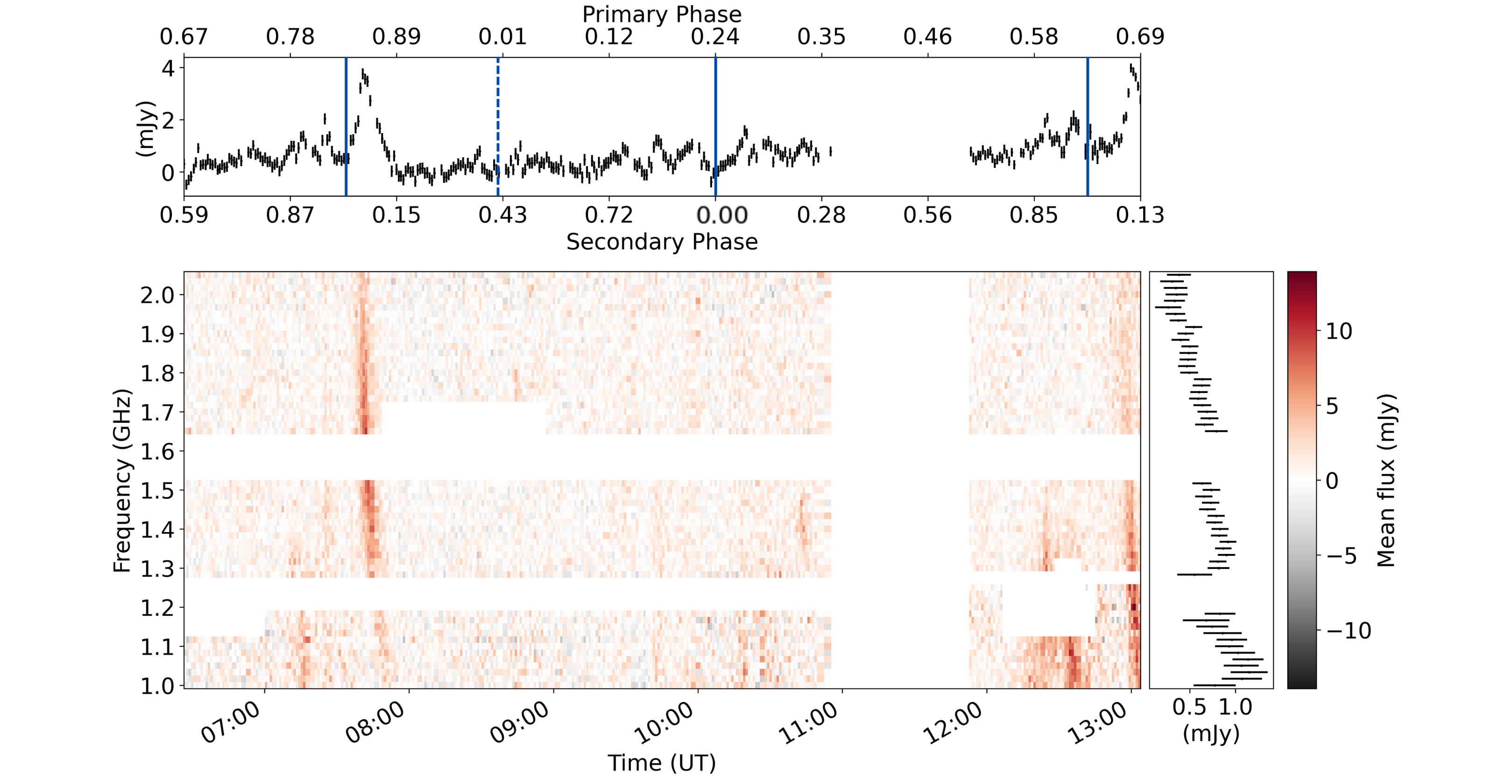}
  \end{subfigure}
\caption{Dynamic spectra for Stokes V for January (top) and February (bottom) observations, taken from visibility plane. The error bars in the light curves and frequency spectrum represent the standard deviation, computed by considering the imaginary component of the amplitude. The top and bottom axis of the top panel represents the primary and secondary phase ($\Phi_1$ and $\Phi_2$), calculated using the primary and secondary period reported in \citep{bouma24}. We fix the reference phase $\Phi = 0$  at the beginning of the first target scan in the January observation. The vertical blue lines indicate the start  ($\Phi=0$) of new primary (dashed) and secondary (solid) cycles, respectively.}
\label{fig:ds_stokes_V}
\end{figure*}

\begin{table*}[h]
 \caption{Properties of the five bursts that have a peak flux at least 4 times the quiescent emission levels of $\sim 0.5$ mJy in Stokes V.}
    \centering
    \begin{tabular}{cccccccccc}
    \hline\hline 
     Central & Duration & Central phase & $\tau_{var}$  & $R^{max}$ & $\log_{10}(T_{b}^{min})$ & $L$ & Drift & V/I & Type \\
     time (UT) & [minutes] & ($\Phi_1,\Phi_2$)& [s]& [$R_\star$]& [K]   & [$10^{23}$ erg s$^{-1}$] & [Mhz/s] & \% \\
     \hline
     21/01-09:50    & $\sim45$ & 0.40, 0.96 & 300 & 179.1 & 6.5 &9.1 & Irregular & $\sim 87$ & F, G? \\
     21/01-12:15 & $\sim30$ & 0.78, 0.93 & 240 & 142.9 & 6.3 &  4.1 & Irregular &$\sim 81$& G? \\
     21/01-13:01 & $\sim10$& 0.88, 0.19& 30 &  16.6  & 8.1 & 3.5 & -1.3& $\sim 88$ & D, G?\\
     20/02-07:40 & $\sim15$ & 0.86, 0.10 & 80 & 47.2 & 7.1 &  2.7 & -0.9 & $\sim 68$ & D?\\
     20/02-13:00 &  $\sim10$ & 0.68, 0.12& 100 & 59.7 & 6.8 & 2.2 & -1.4 & $\sim 76$ & D?\\
    \hline 
    \end{tabular}
    \tablefoot{{The third column indicates the central phases $\Phi_1$ and $\Phi_2$, when the light curve is folded with the primary and secondary period from \cite{bouma24}. The variability timescale $\tau_{var}$ is evaluated as the time it takes for the flux to rise to the peak value, from half of it. This provides the approximate constraints on the maximum size of the emitting region as $R^{max}=c\tau_{var}$, the minimum associated brightness temperature ($T_b^{min}$, see e.g. \cite{smith03}). The last column indicates the tentative labeling of burst type proposed by \cite{bloot24}, relying on the case-by-case evaluation of duration, drift rate, and polarization fraction listed in the purely phenomenological classification of \cite{bloot24}. The luminosity is defined as $L=F_\nu f_b d^2\Delta\nu$ (see text for more details).}}
    \label{tab:flaresl}
\end{table*}

\begin{figure}
\centering
\includegraphics[width=0.47\textwidth]{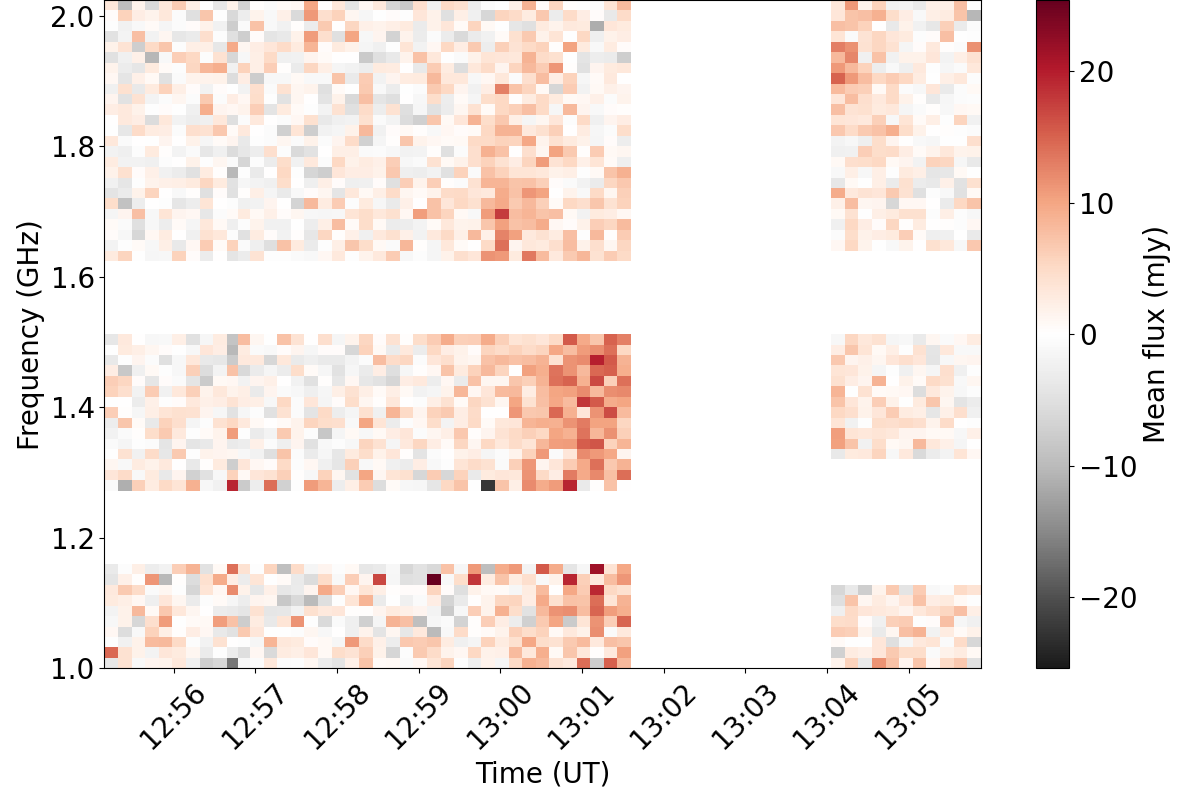}
\includegraphics[width=0.47\textwidth]{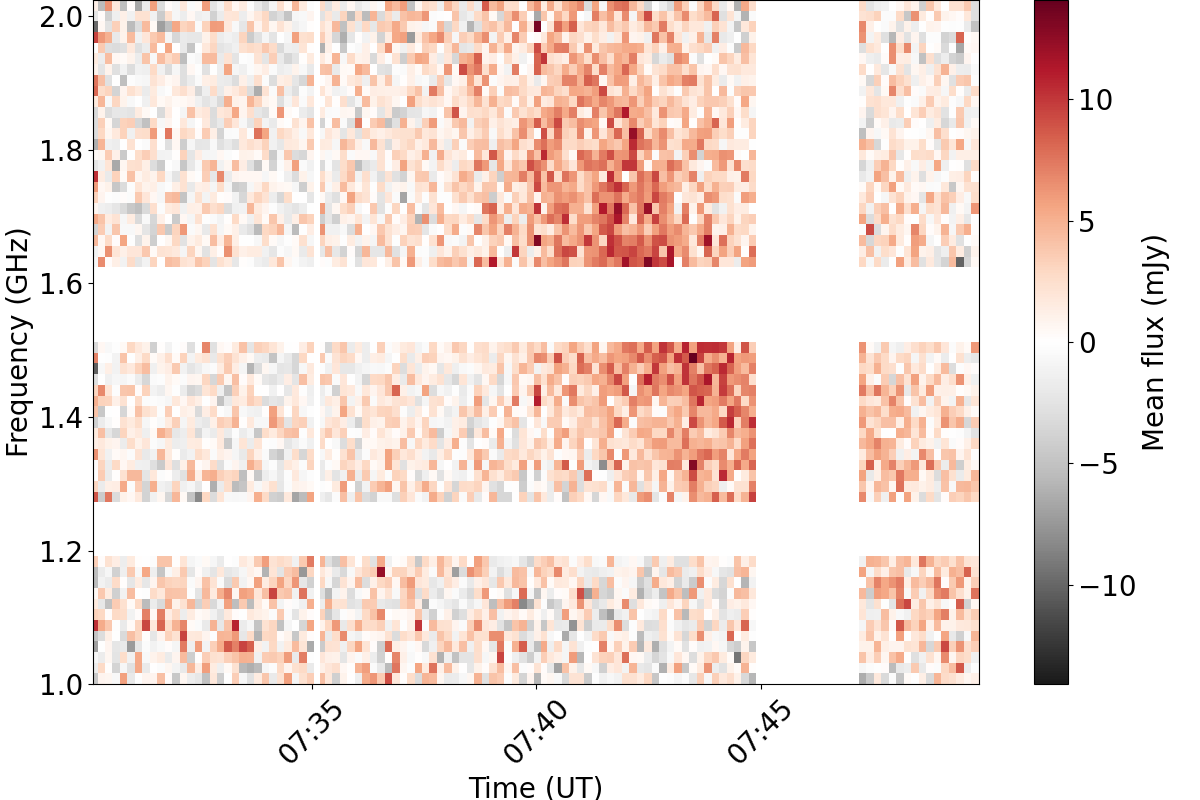}
\includegraphics[width=0.47\textwidth]{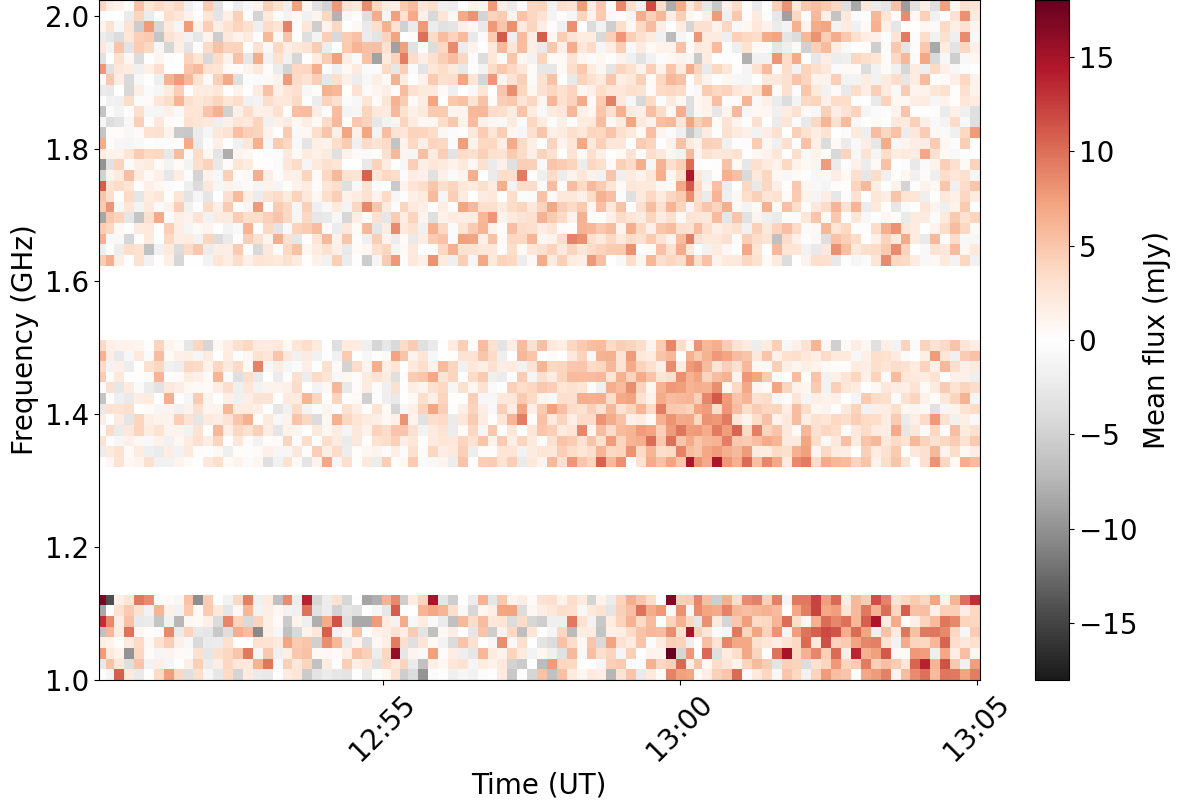}

\caption{Zoomed in version of the dynamic spectra for the three bursts with a clear drift in time and frequency. {\em Left}: third burst in January observation (21/01/18-13:01 UT). {\em Middle}: First burst in February observation (20/02/18-07:40 UT)  in the February observation. {\em Right}: Second burst in February observation (20/02/18-13:00 UT). The dynamic spectra here have been constructed using an integration time of 10 seconds and frequency binning of 16 MHz.}
\label{fig:zoomed_in_DS}
\end{figure}

The VLA L-band emission in both circular components (LL and RR) is detected in the two epochs (see Fig. \ref{fig:carta_map}). The observed radio position is compatible with the previously reported values for DG CVn once we correct for Gaia DR2 proper motion given in Sect. \ref{sec:intro}. Table \ref{tab:flux_statistics} summarizes the main characteristics of the observations and the flux statistics of the January, February and January+February observations. In all cases, the peak and integrated fluxes are compatible with each other, indicating that the source is unresolved. From January to February, the average Stokes I and V fluxes decreases by a factor $\sim 1.5$ and $\sim 3$ in Stokes I and V, respectively. This change is actually related to the presence of bursts with different energetics, rather than a variation of the quiescent emission, as shown in the next subsection.

\subsection{Time variability}

We take advantage of the high S/N ratio of Stokes V emission to investigate variability on short timescales. Figure~\ref{fig:time_series} shows the 5-minute binned light curves, which are phase-folded with the primary and secondary photometric periods, $P_1=6.44$ h and $P_2 = 2.60$ h \citep{bouma24}, and we define the corresponding phases as $\Phi_1$ and $\Phi_2$. Hereafter, we arbitrarily set the reference phase  $\phi_{1,2}=0$ at the beginning of the first target scan of the January observation (21/01/2018-07:17:00 UT). Both in January and February, the LL component remains compatible with being constant and never exceeds the RR component. The latter, instead, varies on different timescales, showing several $\sim 10-45$ minutes-long bursts. During these bursts, the intensity is enhanced up to one order of magnitude, and the net circular polarization fraction can become very high, up to $\sim$90\%. If we consider the total time period during which the source shows an RR excess over the total $11.8$ hours of on-source time, the source appears to have a burst duty cycle of $\sim$15$\%$.

From Fig.~\ref{fig:time_series}, we note that when folded with $P_1$, the emission is quiescent in the phase range $\Phi_1 \simeq 0.90-0.35$, while bursts are scattered and occur at various phases. Unfortunately, the February observations contain a gap precisely at the phases where the main January burst occurred, preventing us from confirming whether bursts repeat around $\Phi_1 \sim 0.4$ in February. On the other hand, when folded with the secondary period $P_2$, the RR excess appears to cluster around the phase range $\Phi_2 \sim 0.8-0.2$.

We refer to App. \ref{app:TESS} for a brief illustration of the TESS light curves, although a dedicated multi-instrument photometric campaign is on-going and will be the subject of a dedicated study with more in-depth discussion. As far as this study of radio behavior is concerned, it is important to note that the optical modulation with $P_2$ is sinusoidal, meaning that the  SSS behavior (i.e., presence of dips) is likely related to $P_1$ only. If confirmed by future observations, and if we assume that the two periods reflect the spins of the two stars, this would suggest that these radio bursts may cluster with the rotational phase of the faster rotator of the binary, and would not be related to the dips, which appear with periodicity $P_1$. On the other hand, it is suggestive to notice that the duration of the most complex bursts is roughly comparable to the duration of the dips.

A possible clustering with the photometric phase has possibly been observed also in J0508-21, the other SSS for which radio light curves have been studied (at 0.55-0.75 GHz, \citealt{Kaur_2024b}). However, in both cases, additional data is required to advance a claim with a sound statistical evidence (like \citealt{bloot24} did for the rotational clustering of radio bursts in AU Mic). Moreover, note that, in our case, we cannot put the TESS and radio light curves in phase, since they are separated by more than 2 years at least, a timescale over which the morphology of the dips is seen to drastically change, and over which the accumulated ephemeris errors make it impossible to maintain the phase coherence (see App.~\ref{app:TESS} for a short discussion).

\subsection{Dynamic spectra}

In Fig.~\ref{fig:ds_stokes_V}, we show the dynamic spectra for Stokes V, generated using the methodology explained in Sect.~\ref{sec:observations}, with a time sampling of 60 seconds and a frequency resolution of 16 MHz. For the January observations, the dynamic spectrum shows the presence of three distinct right-hand circularly polarized (RCP) radio bursts, distributed across different time intervals within the seven-hour observing window. The first burst is the most intense, exhibiting a polarization fraction of almost $90\%$ with the RCP flux reaching over 25 mJy. The analysis of the dynamic spectrum reveals that this burst persists for $\sim$45 minutes. Several fine-scale structures are also evident, including two secondary peaks embedded within the broader burst profile. Specifically, the flux initially increases, followed by a temporary decline, before rising again to form a secondary peak. It subsequently reaches its major peak, followed by a rapid decline into the quiescent phase. During the rising phase of this burst, the emission is predominantly concentrated at the lower frequency range of the L-band, whereas during its peak phase, the flux intensity is more pronounced at the higher end of the L-band.

The second burst of the January observation occurs nearly two hours after the first one and exhibits a lower flux density, with a peak RCP flux of $\sim$15 mJy. This event has a duration of $\sim$30 minutes and is spectrally broad. Throughout its entire duration, the emission is most prominent at the lower end of the L-band. The third burst is detected approximately 30 minutes after the decay of the second burst and is the faintest among the three. It has a relatively short duration, lasting only a few minutes. Unlike the first two bursts, this event is brightest in the central portion of the L-band (between 1.2 and 1.5 GHz) and shows clear evidence of a frequency drift. 

During the first session of the February observations, which lasted 5.5 hours, an RCP flare was detected, with a duration of no more than 10 minutes. When phase-folded with $P_2$ \citep{bouma24}, this flare partially overlaps with the third flare observed in January, occurring around phase $\Phi_2\sim0.1$. This burst also exhibits a clear frequency drift and appears to follow a preceding faint, short-duration event that occurred around 07:10 UT, observable only within the 1–1.2 GHz frequency range.

A similar pattern is observed in the last part of the February session: a short-lived burst, confined to the 1–1.4 GHz frequency range, is detected prior to another broadband burst lasting a few minutes. The latter burst also shows a frequency drift similar to that of the earlier February event.

In order to quantify these frequency drifts, we constructed zoomed-in dynamic spectra for the three short-duration bursts that exhibit negative drifts, as shown in Fig.~\ref{fig:zoomed_in_DS}. These zoomed-in spectra were generated with an integration time of 10 seconds and a frequency bin size of 16 MHz, allowing us to clearly resolve the temporal and spectral evolution of each burst. From these spectra, we manually measured the drift rates, which are reported in Table~\ref{tab:flaresl}, together with other flare properties that we discuss in detail in Sect.~\ref{sec:discussion}. Specifically, the third burst of January shows a drift rate of $\sim -1.3$ MHz s$^{-1}$, while the two February bursts exhibit drift rates of $\sim -0.9$ MHz s$^{-1}$ and $\sim -1.4$ MHz s$^{-1}$, respectively. Notably, when phase-folding the light curve using $P_2$, the three bursts with similar, negative frequency drifts all occur around $\Phi_2\sim0.1$.

Assessing possible frequency drifts in the more structured bursts is more challenging. A closer look at the brightest burst (the first in the January observation) hints for instance at the presence of a positive frequency drift at the third sub-structure, but the presence of overlapping components with possibly different drifts, and the unavailability of flagged channels prevent us from a detailed evaluation.

\begin{figure}[t]
\centering
\includegraphics[width=0.49\textwidth]{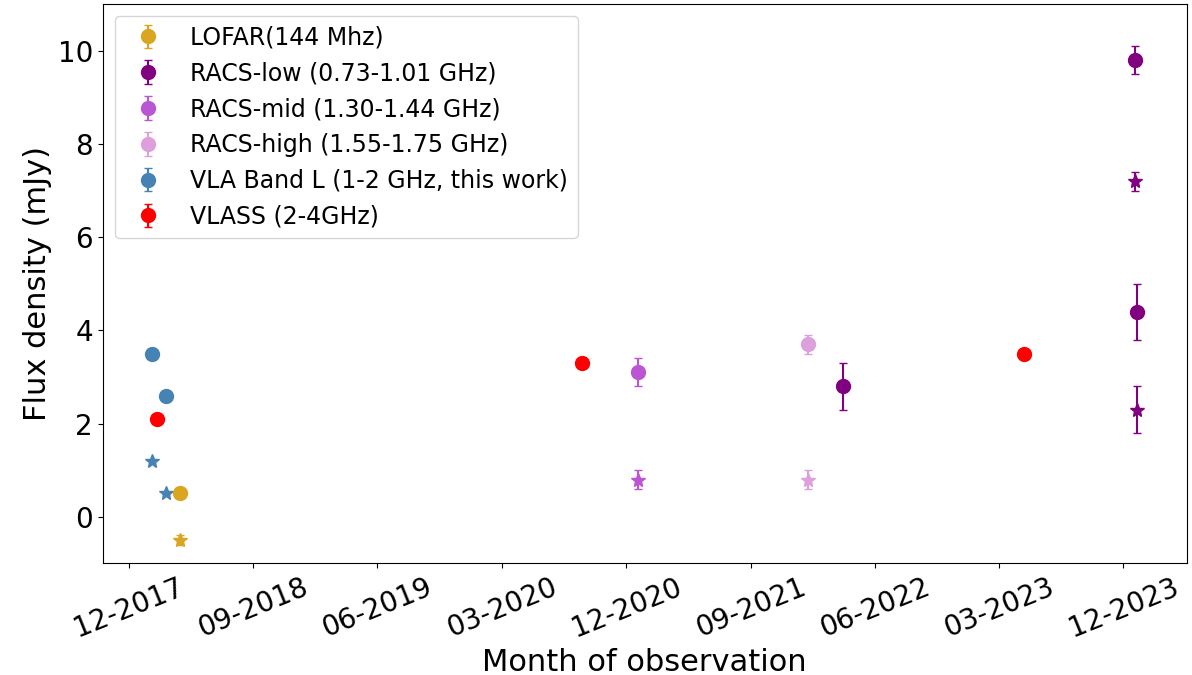}
\caption{Available radio detections of DG CVn, including the VLA band L 2018 observations analyzed in this work, and the archival data from LOFAR, ASKAP and VLA S-band surveys, as collected in the Radio Stars catalogue, \url{radiostars.org} \citep{driessen24}. Colors identify different surveys/observations, with circles representing Stokes I and stars representing Stokes V detections (if available), respectively.}
\label{fig:longterm_lc}
\end{figure}

\subsection{Information from other archival radio datasets}\label{sec:long term radio light curve}
In addition to the band L observations that are the focus of this study, we also present here the other available information on DG CVn radio emission at different wavelengths in Fig. \ref{fig:longterm_lc}, as obtained from various archival observations over the years. The data points include: (i) the V-LoTSS \citep{callingham23} observations at 144 MHz, where the source was detected with a $\sim$100\% left circularly polarized emission of $0.5\pm0.1$ mJy; (ii.) the VLASS survey (S band, 2--4\,GHz), with a Stokes I flux density of $2.1 \pm 0.1$ mJy, $3.3 \pm 0.1$ mJy and $3.5 \pm 0.1$ mJy in epochs 1.1, 2.1 and 3.1, respectively; (iii.) the all-sky Rapid ASKAP Continuum Survey (\citealt{driessen24}) at 0.887 GHz (RACS-low; \citealt{mcconnell20,hale21}), 1.367 GHz (RACS-mid; \citealt{duchesne23}), and 1.655 GHz (RACS-high; \citealt{Duchesne25}), with flux density in Stokes I ranging between $\sim$3-5\,mJy in three pointings, while it reaches $\sim$ 10 mJy during one of the two pointings of the RACS-low survey. We have also included the Stokes V fluxes, wherever the S/N>4. This gives the circular polarization fraction varying between 25-80$\%$ for these pointings at respective ASKAP bands.

Note also the apparent long-term variability of the values are likely contaminated by the presence of bursts: the clear example is from the data of this work, for which the time-integrated flux for January and February is mostly due to the different peak values of the main bursts, rather than a different quiescent emission (which looks similar in both observations: $\sim$3 and $\sim$0.5 mJy in Stokes I and V, respectively, see Fig. \ref{fig:time_series}).

\section{Discussion}\label{sec:discussion}

The analysis of these two long observations of DG CVn shows that there is an apparently quiescent, weakly polarized component, which could be ascribable to gyrosynchrotron, incoherent emission coming from the magnetosphere surrounding one or both stars, even though the observed radio luminosity is orders of magnitude brighter than that predicted by G\"udel-Benz relation, similar to what has also been observed for UV Cet \citep{bastian22}.  More importantly, we have detected multiple radio bursts, ranging from short duration bursts, lasting for $\lesssim 10$ minutes, to the ones that evolve for $\gtrsim 30$ minutes. As can be seen in Fig. \ref{fig:time_series}, all bursts are possibly fully RCP (no LL excess is seen). 

In Table \ref{tab:flaresl}, we summarize the characteristics of the bursts, indicating the range of primary and secondary phases encompassed, frequency, drift rate and a tentative match to the phenomenological classification of M dwarf bursts proposed by \cite{bloot24}. We also report the luminosity, $L = 4\pi d^2 F_\nu f_{b}\Delta\nu$, where we adopt $\Delta\nu = 1$ GHz (the observed bandwidth) and assume a beaming fraction $f_b = 1.6$ sr, representative of the values for Jupiter's auroral oval (e.g., \citealt{griessmeier07}). Additionally, we estimate the maximum size of the emitting region as $R^{\max} = c\tau_{var}$, where the variability timescale $\tau_{var}$ is measured from the flux rise time of each burst.
These emitting radii are upper limits, and are tens to hundred times larger than the stellar radius, $R_{\star}=0.72
\,R_\odot$ \citep{bouma24}. Such values of peak fluxes and $R^{max}$ imply lower limits on the brightness temperature $T_b \gtrsim 10^{6}-10^8$ K. However, these limits are not constraining enough to formally provide a conclusion on the emission mechanism, but, combined with the high circular polarization, they clearly points towards coherent plasma or ECM emission. As explained in detail by \cite{villadsen19}, it is not trivial to practically distinguish these mechanisms for a given burst in M dwarfs. In the solar system paradigm, plasma emission is typically associated with stochastic solar flares and associated coronal mass ejections (CMEs), while ECM is often associated with aurora such as on Jupiter and possibly on the Sun \citep{Yu}. In M dwarfs, despite the high flaring rates in many cases, a definitive confirmation of CMEs (i.e., a huge amount of material moving outward at very high speed) is still pending \citep{Crosley_osten_2018,zic20,Mohan_2024}, and the much higher large-scale magnetic fields may indeed prevent them \citep{Alvarado_gomez_2018, villadsen19}.

Considering these caveats, there are two main ways to distinguish the two mechanisms. The first one is to infer the velocity associated to the frequency drift rate. Both ECM and plasma emission can show drift in frequency over time, that could be caused due to beaming effects or the motion of the source. If $H_{\nu}$ is the scale height for the peak emission frequency $\nu$ (where $H_\nu$ is defined as the plasma density scale height for plasma emission, or the magnetic field lengthscale for ECM), the velocity of the moving source ($V$) can be calculated as \citep{villadsen19}:
\begin{equation}
V = H_{\nu}\left(\frac{\dot\nu}{\nu}\right)\, .
\end{equation}
Taking the negative drift rates of Table \ref{tab:flaresl}, and assuming $H_{\nu}\sim 0.1$-$1 \,R_\star$ (which covers reasonable values, see \citealt{villadsen19}), we infer a range of velocities $V \sim 30$-$600$ km/s. This range of values is compatible with expanding coronal loops and is much lower than those expected from the forward shock of a CME, which must travel at super-Alfvénic speeds and can reach up to several thousands of kilometers per second in case of magnetized ($B\sim$ kG) M dwarfs \citep{villadsen19}. However, the drift could also be due to a beaming effect, which is well-documented and modeled for ECM \citep{Hess_2008} and for which we see emission coming from regions with different magnetic field strengths and/or plasma densities, as the star rotates. 

The second method to infer the underlying mechanism is to  observe a rotational modulation, which would favor ECM, since it would be related to specific magnetic field lines which cross the line of sight once or twice per period. This would again indicate the presence of the angular beaming characteristic of ECM as this kind of emission is primarily beamed perpendicular to the local magnetic field lines, forming a hollow-cone pattern around them and can only be detected when this beam aligns with our view. As a matter of fact, bursts in AU Mic and other M dwarfs \citep{bloot24} tend to be clustered in two sections of the stellar rotational phase. In our data, the short duration bursts ($\lesssim 10$ minutes long with a similar frequency drift pattern, i.e. the third burst in January observation and second and fourth burst in February observation) occur around a (secondary) phase range of $\Phi_2 \sim 0.9$-$0.2$, while no clear clustering is seen when the light curves are folded using the primary period.  In particular, we notice the similarity in shape between the main complex bursts in January and February, happening at around $\Phi_2 \sim 0.8$-$0.15$. If the phase bunching is confirmed, this might trace the magnetic field and plasma density characteristics of the emitting region, associated with the detectable beamed emission directed toward our line of sight at that phase range.

Note that we see clustering of RCP bursts only whereas AU Mic has both the LCP and RCP counterparts, but this could be a geometrical effect, which allows the observer to intersect one or both magnetic poles, as invoked to explain UV Cet's RCP-only bursts based on its magnetic map \citep{bastian22}. Bottom-line, given the scarcity of data, we cannot put statistically meaningful constraints on the geometry as done for AU Mic by \cite{bloot24}. However, we can argue that, assuming that ECM is the mechanism underlying the coherent, polarized bursts, the local magnetic field in the emitting regions is of the same polarity, and of the order of $\sim$ 3-5 kG, although this is biased by the data availability limited to the L-band only.

We also compared the bursts observed on DG CVn with the different phenomenological classes of bursts seen on AU Mic \citep{bloot24}. Our observed DG CVn bursts can be tentatively assigned to type D (high drift rates, 30-60 minutes), F (irregular) and G (broadband, short timescales), see Table~\ref{tab:flaresl}. Note that such a case-by-case manual classification is just meant to draw possible analogies with other M dwarfs, but can present ambiguities and has no strong physical implications. While different frequency/temporal resolution makes detailed comparison difficult, some similarities suggest that DG CVn shares the properties with other radio-bursting M dwarfs \citep{zhang23,bloot24,pineda23,bastian22}. For example, the tentative presence of G-type bursts places DG CVn in closer similarity to UV Cet, whereas no such bursts have been observed in AU Mic. This indicates that DG CVn emission properties may resemble those of UV Cet.

Additionally, the known binarity and SSS behavior of this source provide an interesting context, on which the radio emission can offer additional insights.
In this regard, we remark the striking similarity of the radio light curves and optical photometric periodicity between DG CVn and J0508-21 \citep{Kaur_2024b}, which are the only two SSSs studied in detail in radio so far. In both these systems, the RR excess is seen at a specific phase \citep{Kaur_2024b}. The periods of SSSs are stable over a timescale of a few years at least. However, the morphology of the light curve changes. The depths of the dips and the phases at which the dips occur are stable over $\sim$ tens to hundreds of cycles, but rarely thousands of cycles. Fig. 6 of \cite{bouma24} shows that none of $\sim$ 30 SSSs analyzed in their paper with a > 2-year baseline maintained a similar light curve morphology, although they usually remained complex. Because of this, and due to the loss of phase coherence over the $\gtrsim 2$ yr gap between the closest TESS and VLA observations, we cannot phase-connect the optical and radio light curves of DG CVn, as \cite{Kaur_2024b} did for J0508-21, thanks to a much smaller time gap (few months). This implies that we don't have the information needed to investigate possible correlations in phase between the radio bursts and the optical dips.

Nevertheless, if future observations confirm the bursts to be periodic on $P_{2}$, it would mean that the observed radio behavior is reflecting the radio loudness of the fast rotator in the binary, and is not related to the SSS behavior which is seen only with a periodicity $P_{1}$. This would suggest that the coherent bursts are more directly linked to stellar youth and rapid rotation, as in the case of AU Mic, rather than being caused by transiting material at the co-rotation radius. We highlight the need for simultaneous optical and radio observations to have multiple cycles close in time, with minimum evolution of the obscuring material. Ultimately, simultaneous multi-wavelength monitoring will be key to uncovering the true driver of the coherent bursts, whether rooted in stellar magnetism, rotation, or circumstellar dynamics.

\begin{acknowledgements}
SK carried out this work within the framework of the doctoral program in Physics of the Universitat Aut\`onoma de Barcelona. SK, OM, and DV are supported by the European Research Council (ERC) under the European Union’s Horizon 2020 research and innovation programme (ERC Starting Grant "IMAGINE" No. 948582).  E.I. acknowledges funding from the European Research Council under the European Union's Horizon Europe programme (grant number 101042416 STORMCHASER). We acknowledge financial support from the Agencia Estatal de Investigaci\'on (AEI/10.13039/501100011033) of the Ministerio de Ciencia e Innovaci\'on and the ERDF ``A way of making Europe'' through projects
  PID2023-146675NB-100,         % JMG
  PID2022-137241NB-C4[1:4],    % CAB+IAA+IAC+UCM
  PID2021-125627OB-C31,        % ICE
  PID2020-117710GB-100,         % JMG
  PID2020-117404GB-C21,         % MPT
  PID2023-147883NB-C21,          %MPT
  %RYC2021-032892-I,             % ASM
% Add here any additional PID/RYC project
and the Centre of Excellence ``Severo Ochoa''/``Mar\'ia de Maeztu'' awards to the Institut de Ci\`encies de l'Espai (CEX2020-001058-M), Instituto de Astrof\'isica de Canarias (CEX2019-000920-S), and Instituto de Astrof\'isica de Andaluc\'ia (CEX2021-001131-S).
The National Radio Astronomy Observatory and Green Bank Observatory are facilities of the U.S.~National Science Foundation operated under cooperative agreement by Associated Universities, Inc.

\end{acknowledgements}

\bibliographystyle{aa}
\bibliography{output}

\onecolumn

\appendix

\section{TESS light curves}\label{app:TESS}

\begin{figure}[h]
  \centering

  \vspace{-0.2cm}
  \parbox{0.33\textwidth}{\centering Sector 23}%
  \parbox{0.33\textwidth}{\centering  Sector 50}%
  \parbox{0.33\textwidth}{\centering  Sector 77}

  % Row 1
  \begin{subfigure}{0.33\textwidth}
    \includegraphics[width=\linewidth]{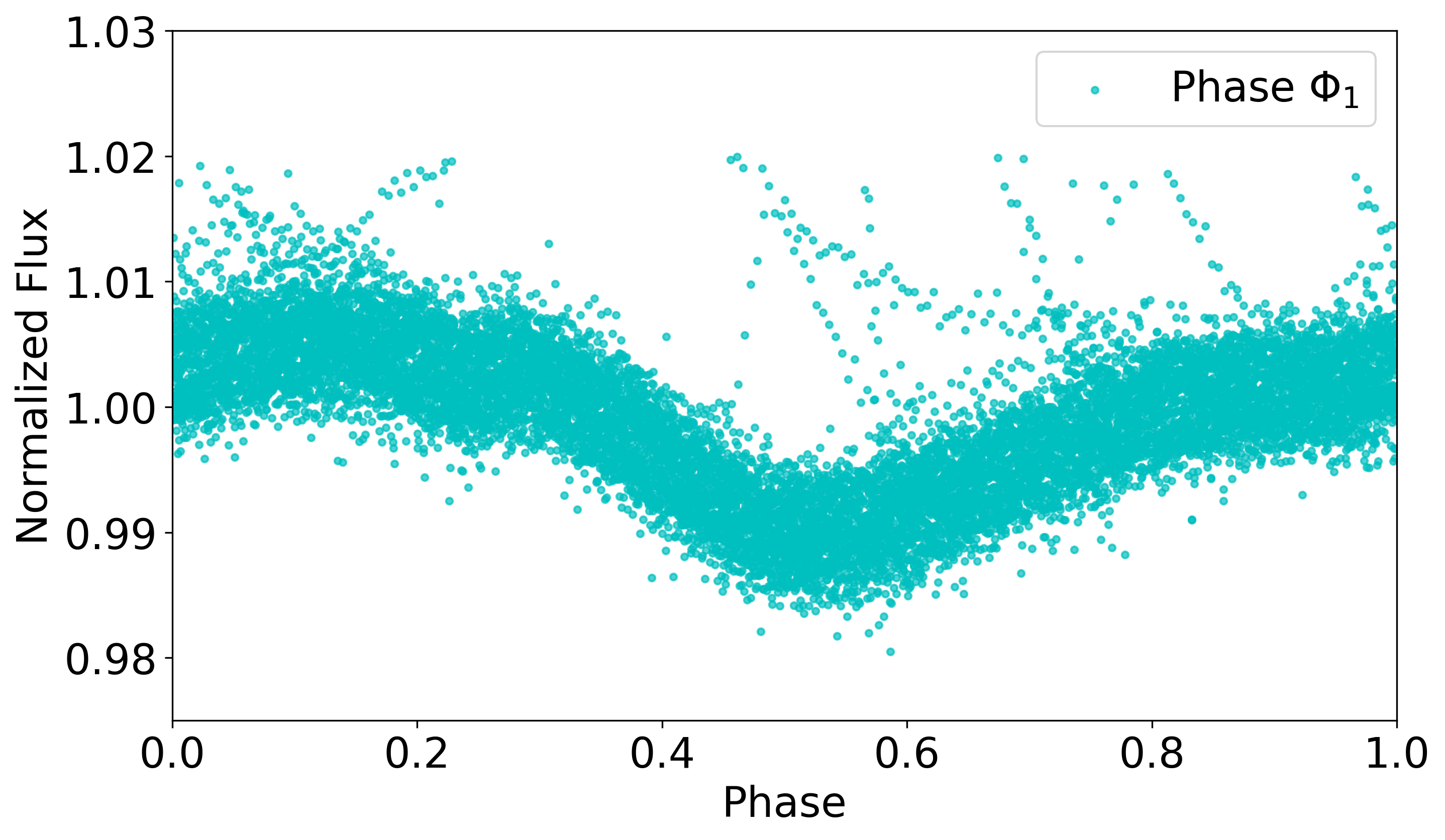}
  \end{subfigure}
  \hfill
  \begin{subfigure}{0.33\textwidth}
    \includegraphics[width=\linewidth]{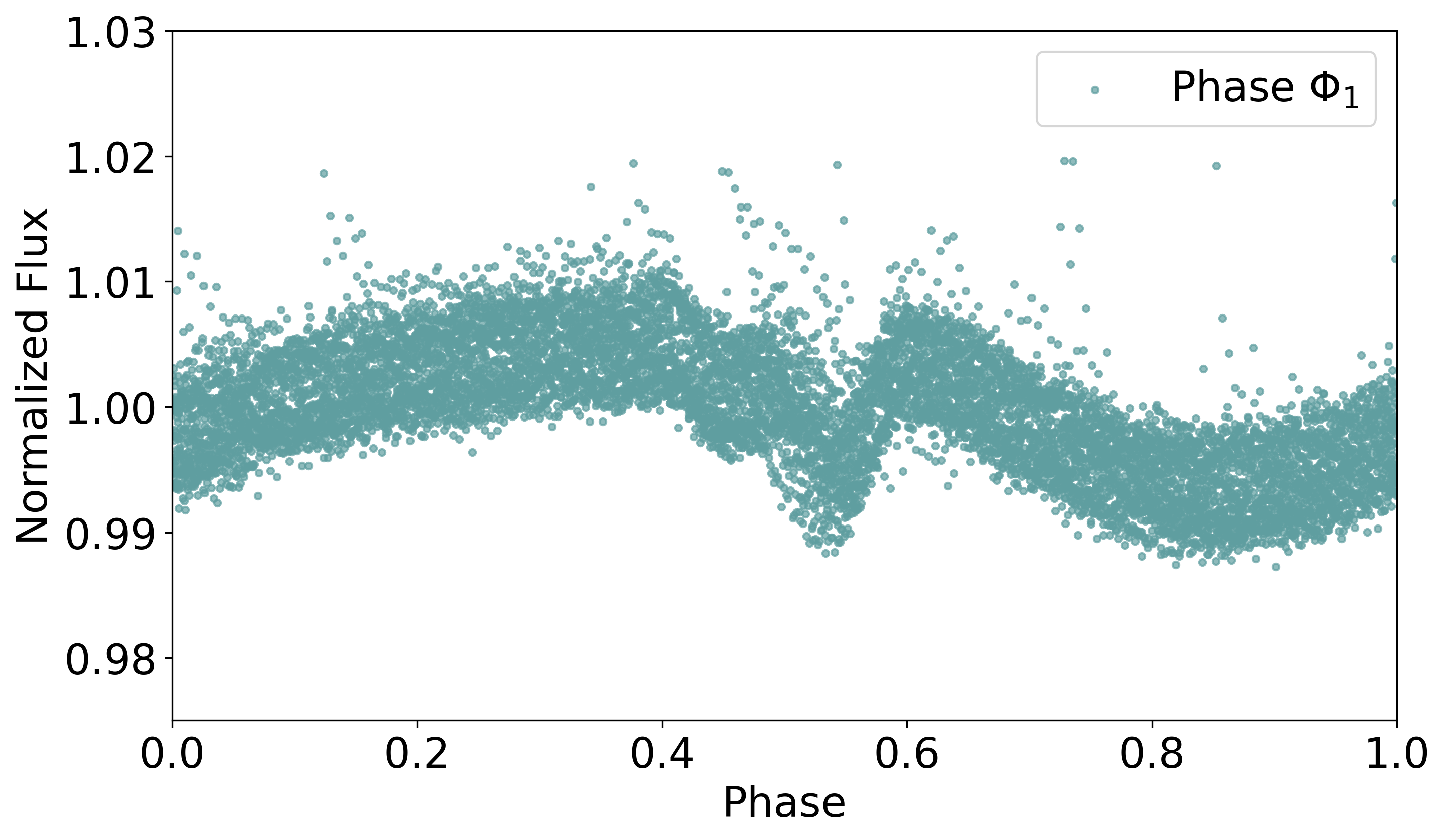}
  \end{subfigure}
  \hfill
  \begin{subfigure}{0.33\textwidth}
    \includegraphics[width=\linewidth]{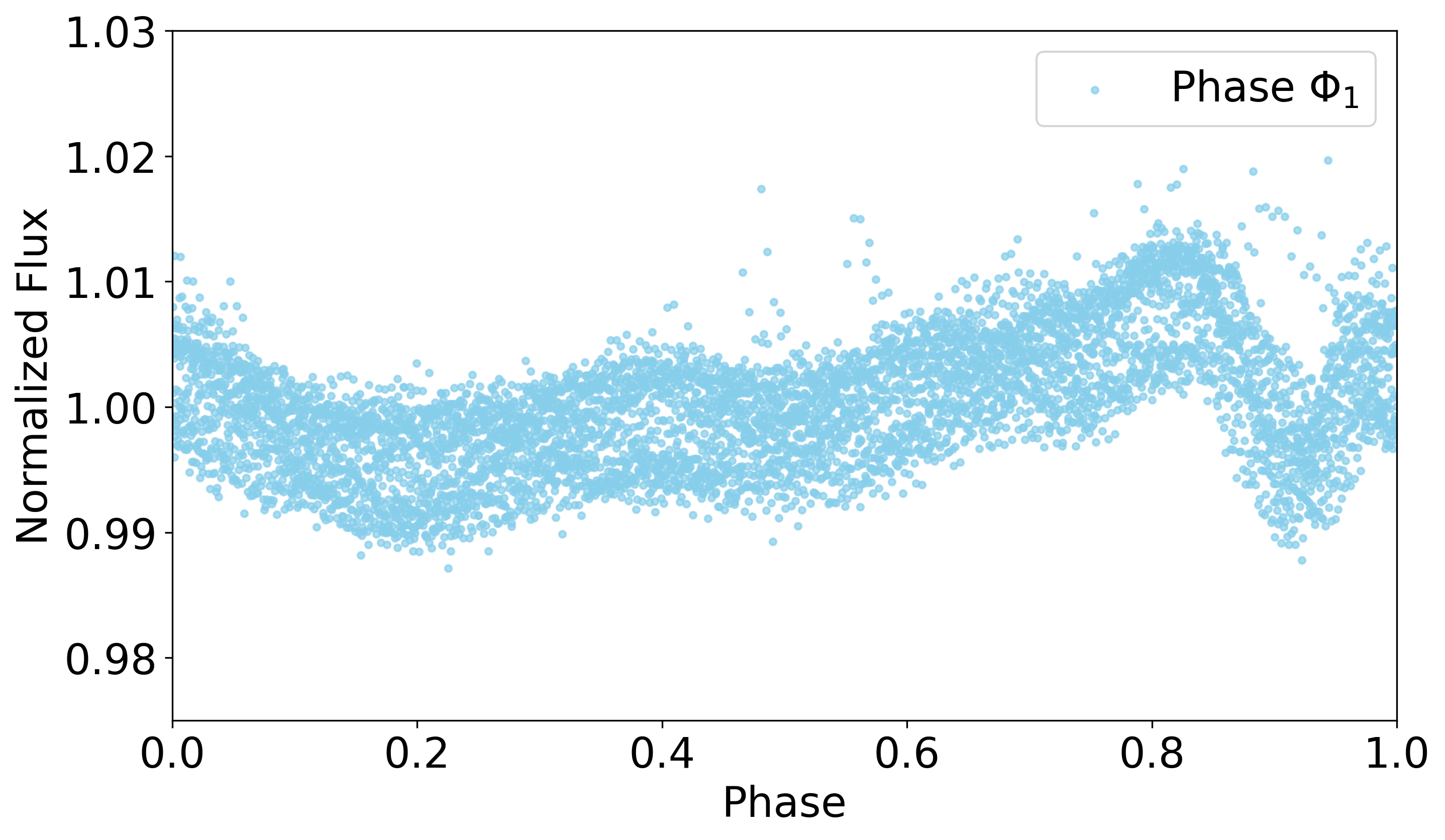}
  \end{subfigure}

  \vspace{0.5cm}

  % Row 2
  \begin{subfigure}{0.33\textwidth}
    \includegraphics[width=\linewidth]{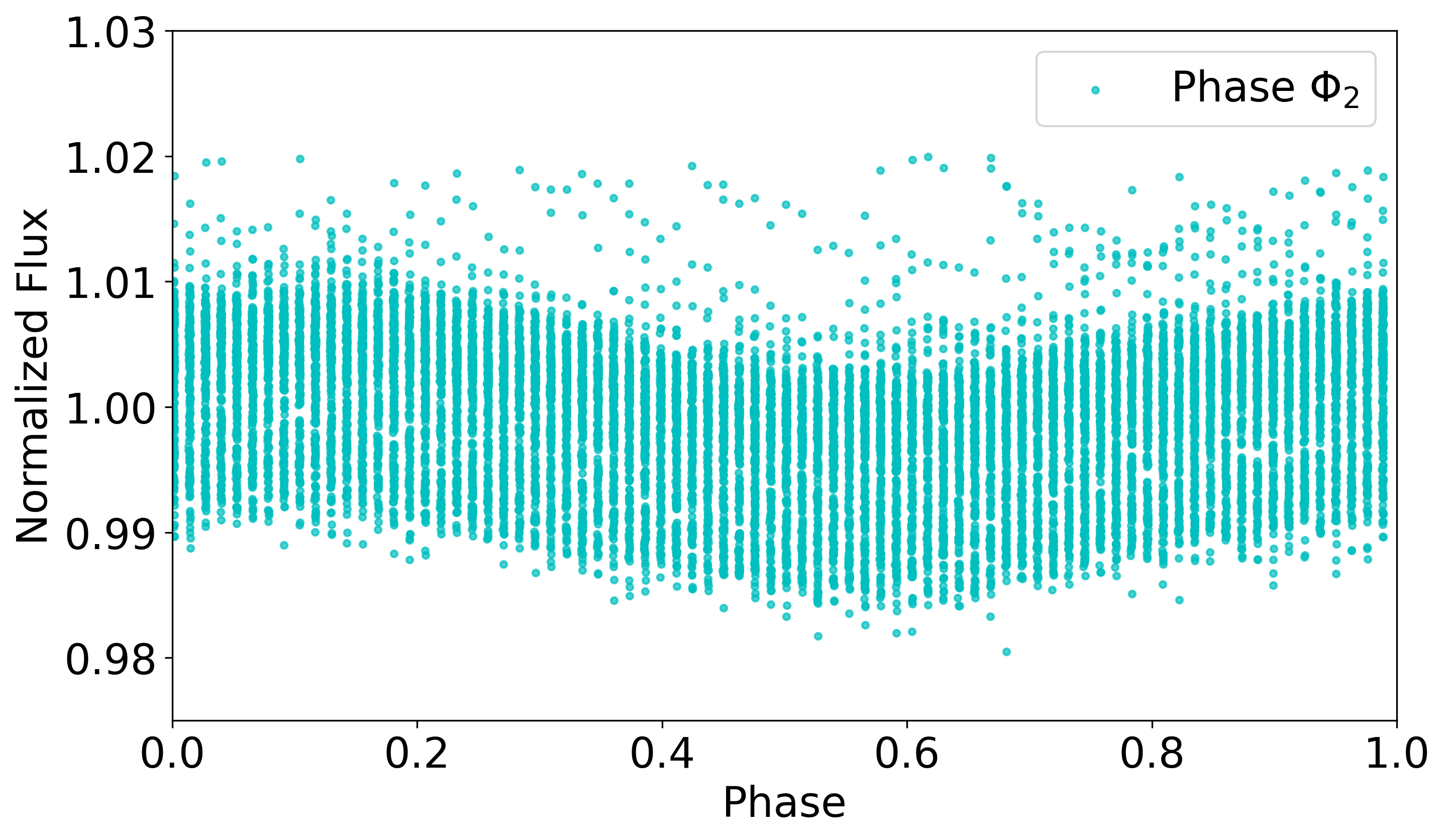}
  \end{subfigure}
  \hfill
  \begin{subfigure}{0.33\textwidth}
    \includegraphics[width=\linewidth]{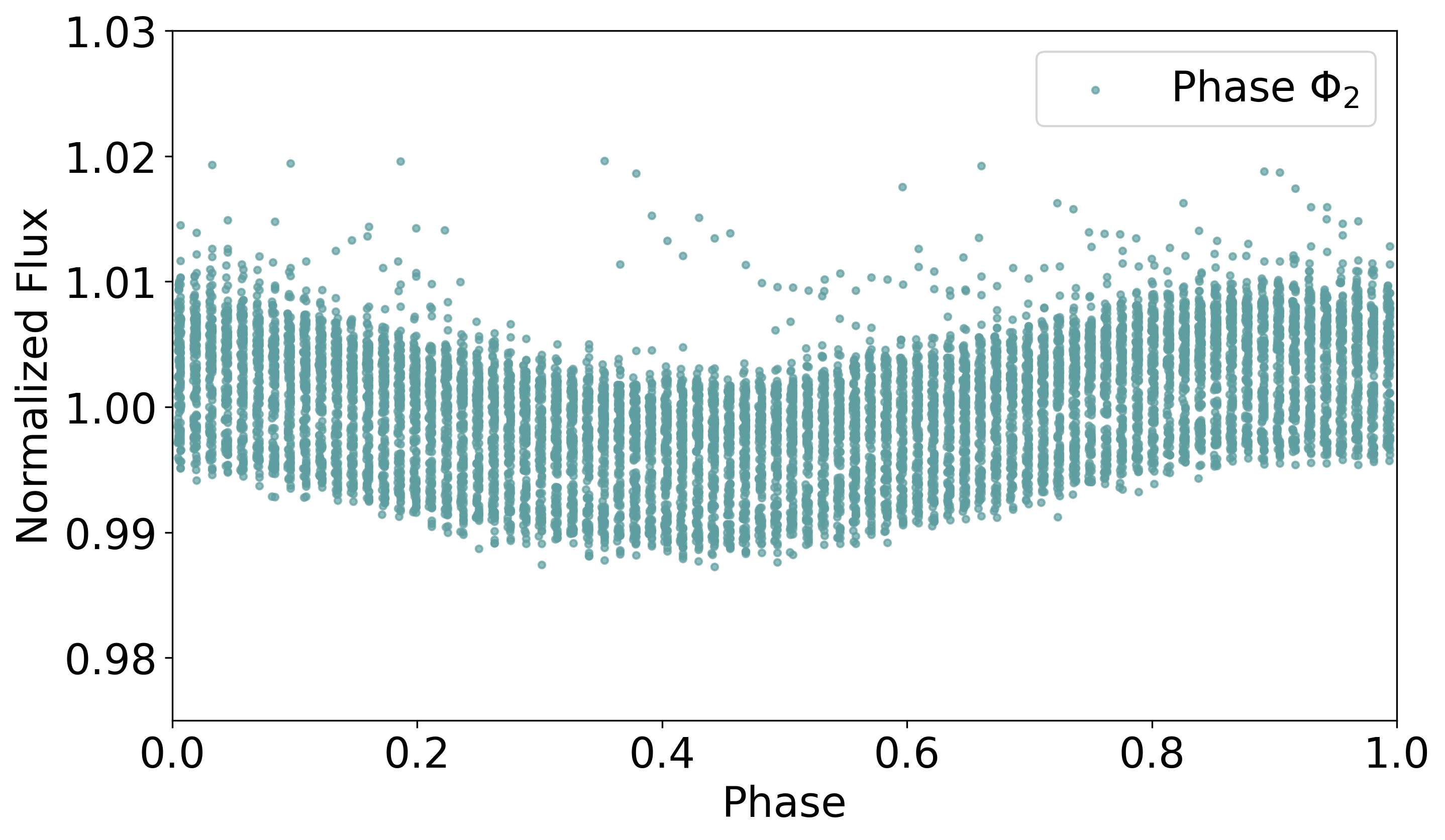}
  \end{subfigure}
  \hfill
  \begin{subfigure}{0.33\textwidth}
    \includegraphics[width=\linewidth]{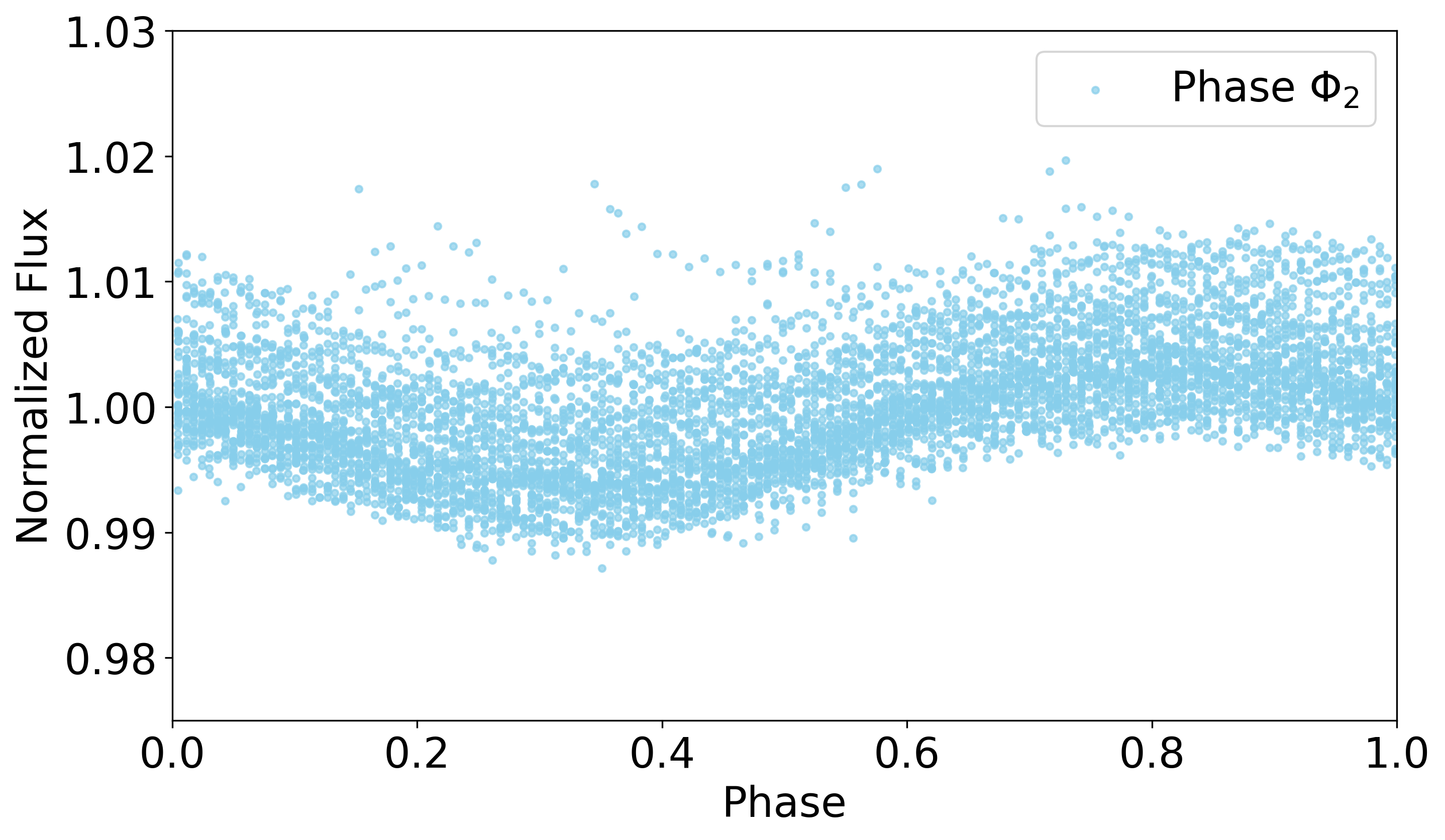}
  \end{subfigure}

  \vspace{0.5cm}

  % Row 3
  \begin{subfigure}{0.33\textwidth}
    \includegraphics[width=\linewidth]{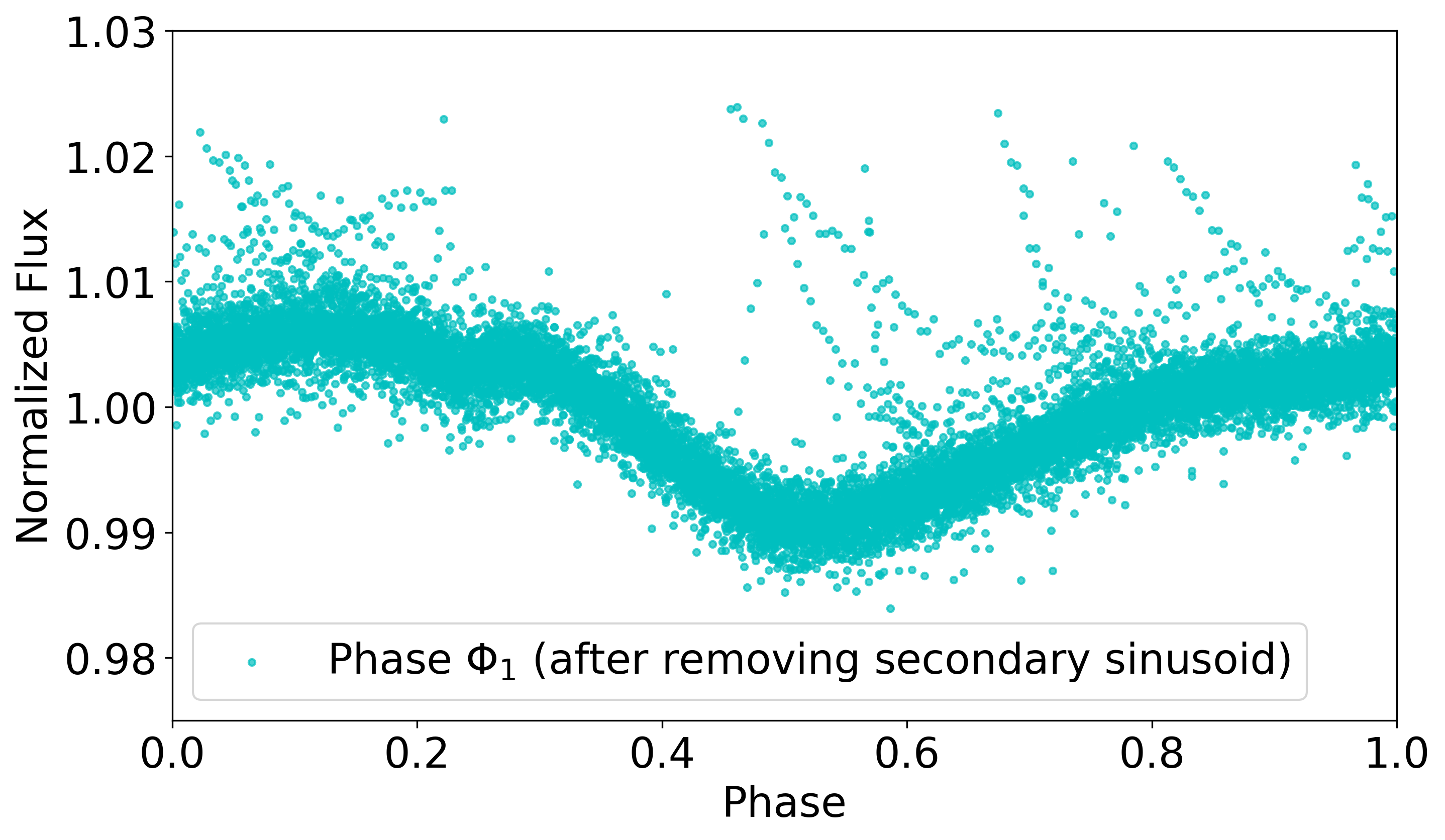}
  \end{subfigure}
  \hfill
  \begin{subfigure}{0.33\textwidth}
    \includegraphics[width=\linewidth]{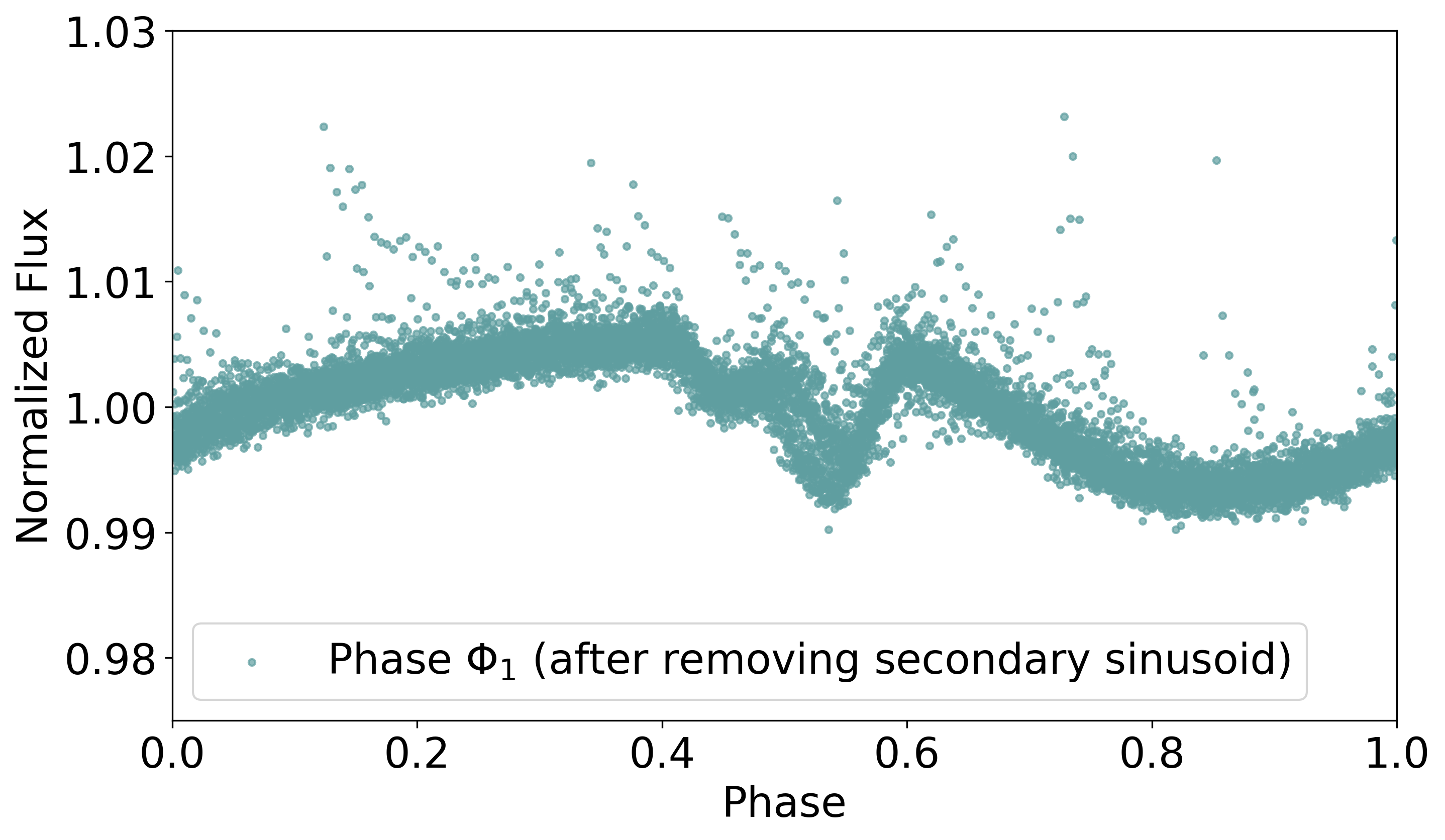}
  \end{subfigure}
  \hfill
  \begin{subfigure}{0.33\textwidth}
    \includegraphics[width=\linewidth]{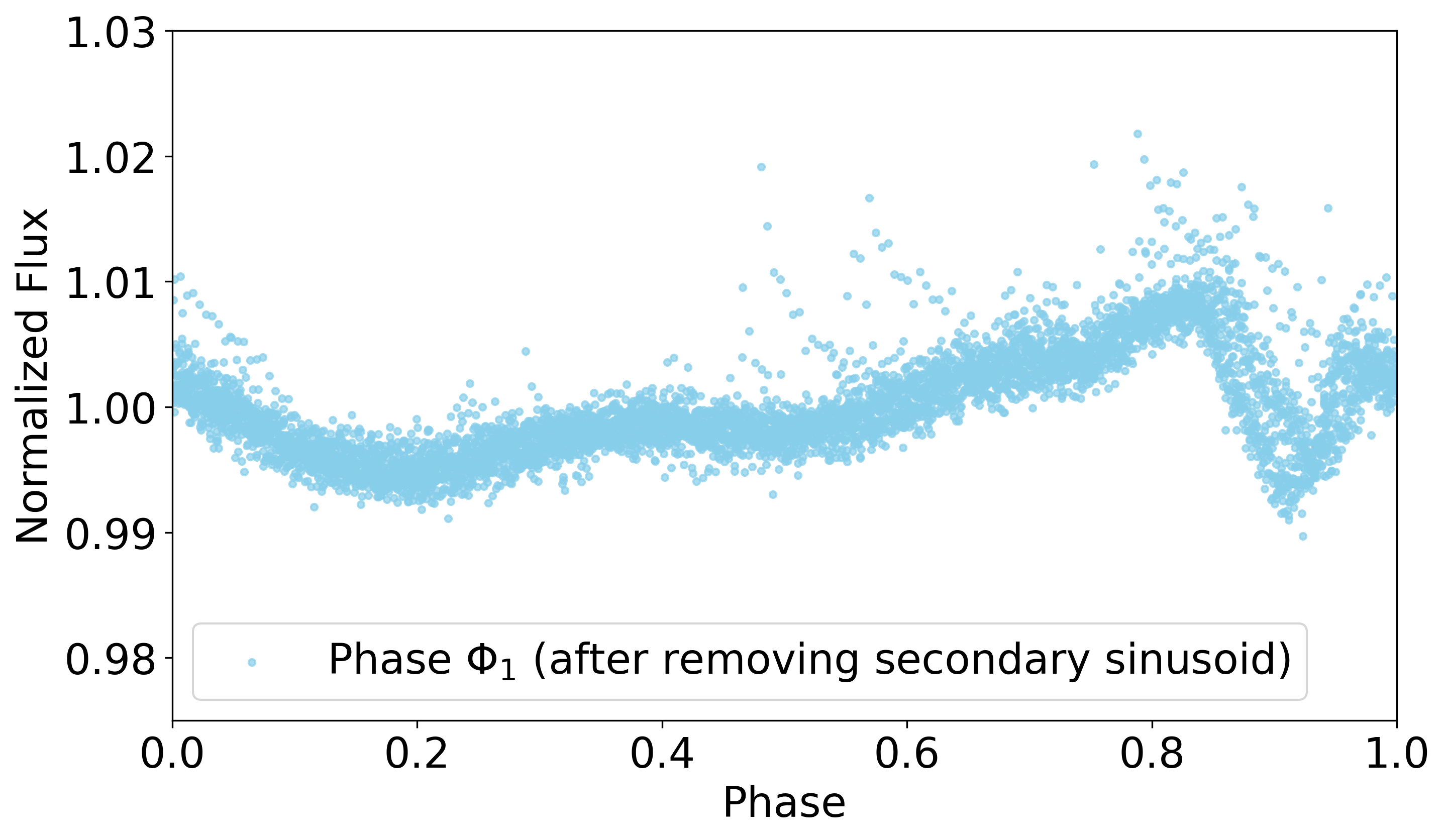}
  \end{subfigure}

  \caption{ Simple Aperture Photometry flux light curves of TESS sectors 23 (left column), 50 (middle column) and 77 (right column), with a cadence of 2 minutes. The top row represents the normalized flux phase folded with the primary period of $P_1 = 6.44$ hours while the middle row represents the same normalized flux phase folded with the secondary period of $P_2 = 2.60$ hours. In the third row, we present the flux folded with the primary period, but after removing the secondary signal. In all these  light curves, the phase is arbitrarily set by fixing $\Phi = 0$ at the beginning of the first target scan of VLA January observation (21/01/2018-07:17:00 UT), as in the rest of the paper.}
  \label{fig:TESS_lc}
\end{figure}

We present the phase-folded 2-minutes cadence TESS light curves of DG CVn, aka TIC 368129164, in Fig. \ref{fig:TESS_lc}. Compared to \cite{bouma24} where sectors 23 (March-April 2020, left column) and 50 (March-April 2022, middle column) have been reported, here we add a third available TESS sector 77 (March-April 2024), and, besides folding the light curves with $P_1$ (top panels), we focus also on the modulation with the second period, $P_2$ (second row panels). The third row shows the folding with $P_1$ after removing the $P_2$ signal, which clearly decreases the dispersion of the points at a given phase.

If we interpret the periods as the rotational spins of the two stars, this implies that the SSS behavior is associated to the slowest of the two components only. Moreover, as in many other SSSs, within a given sector the light curve show consistently repeating features (and frequent, stochastic flares), but from one sector to another the dips shift in phase and shape. Indeed, dips are almost invisible in sector 23, only tentatively seen around $\phi_1=0.25$ (indeed dubbed as ambiguous by \citealt{bouma24}). The change in shape is likely indicative of the variability of the optical depth and/or size of the obscuring material. The shift in phase could be one or more of the following reasons: (i) intrinsic, due a non-perfectly co-rotating orbit or, equivalently, a shift in the phase position of the obscuring material or the magnetic field lines that trap the material; (ii) indicative of the lifetime of opaque clumps (related to disk debris and/or disintegrating planet), forming and fading away at different phases over the years; (iii) importantly, caused by the loss of phase coherence after typically ${\cal O}$(yr), due to the uncertainty on the dip phase, which eq. (9) in \cite{bouma24} evaluated as $\delta\phi\sim 0.02~\Delta t/(20~{\rm d})$ after a period $\Delta t$ (specifically,  $\delta\phi\sim 0.8$, i.e. a complete loss of phase coherence, considering the $\Delta t\sim 2.2$ yr between the latest VLA and the earliest TESS observations)..

With TESS data only, we cannot favor any of the three explanations above, and more discussion about this issues, common in SSSs, can be found in \cite{bouma24}. In this sense, more details of a dedicated, on-going, multi-instrument photometric campaign will be the object of a dedicated study, while this article is focused on the radio emission alone.\\

 \vspace{0.5cm}

\end{document}